\documentclass[12pt]{article}

\usepackage{graphicx}
\usepackage{xspace}
\usepackage{amsmath}
\usepackage{amsfonts}
\usepackage{amssymb}
\usepackage{mathrsfs}

\usepackage[colorlinks,linkcolor=black,citecolor=blue,urlcolor=blue,linktocpage]{hyperref}
\newcommand{\hhref}[2][]{\href{http://arxiv.org/abs/#2#1}{arXiv:#2}}

\makeatletter
\def\section{\@startsection {section}{1}{\z@}{-3.5ex plus -1ex minus -.2ex}{2.3ex plus .2ex}{\large\bf}}
\def\subsection{\@startsection{subsection}{2}{\z@}{-3.25ex plus -1ex minus -.2ex}{1.5ex plus .2ex}{\normalsize\bf}}
\makeatother
\makeatletter

\@addtoreset{equation}{section}

\makeatother

\def\patricia{Centro Brasileiro de Pesquisas F\'isicas - CBPF, 22290-180 Rio de Janeiro, Brazil}
\def\patemail{\footnote{Speaker, patricia.rebello.teles@cern.ch}}

\def\david{CERN, EP Department 1211 Geneva, Switzerland}
\def\davidemail{\footnote{dde@cern.ch}}

\def\daniel{Universidade Federal do Rio de Janeiro - UFRJ, 21941-901 Rio de Janeiro, Brazil}
\def\danielemail{\footnote{dan.ernani@gmail.com}}

\def\Title#1{\begin{center} {\Large #1 } \end{center}}
\def\Author#1{\begin{center}{ \sc #1} \end{center}}
\def\Address#1{\begin{center}{ \it #1} \end{center}}

\newenvironment{Abstract}{\begin{quotation}  }{\end{quotation}}
\newenvironment{Presented}{\begin{quotation} \begin{center} 
             PRESENTED AT\end{center}\bigskip 
      \begin{center}\begin{large}}{\end{large}\end{center} \end{quotation}}

\def\beq{\begin{equation}}
\def\eeq#1{\label{#1}\end{equation}}
\def\eeqn{\end{equation}}

\def\beqa{\begin{eqnarray}}
\def\eeqa#1{\label{#1}\end{eqnarray}}
\def\eeqan{\end{eqnarray}}

\let\bar=\overbar

\def\etal{{\it et al.}}
\def\ie{{\it i.e.}}

\def\Dslash{\not{\hbox{\kern-4pt $D$}}}
\def\dslash{\not{\hbox{\kern-2pt $\del$}}}

\def\ee{e^+e^-}

\def\msb{{\bar{\ssstyle M \kern -1pt S}}}

\newcommand{\bbbar}     {\ensuremath{b\bar{b}}}
\newcommand{\Lumi}{\mathcal{L}}
\newcommand{\LumiInt}{\mathcal{L}_{\rm \tiny{int}}}
\providecommand{\gaga}{{\gamma\gamma}}
\newcommand{\sqrts}{\sqrt{s}}
\newcommand*{\cm}{c.m.\@\xspace}
\providecommand{\pythia}{{\sc pythia}}
\providecommand{\superchic}{{\sc superchic}}
\providecommand{\madgraph}{{\sc madgraph}}
\providecommand{\fastjet}{{\sc fastjet}}
\def\ttt#1{\texttt{\small #1}}

\begin{document}

\setcounter{page}{0}
\thispagestyle{empty}

\begin{titlepage}

\vfill
\Title{Measurements of $\gaga \to$~Higgs and $\rm \gaga \to W^{+}W^{-}$ \\
in $\ee$ collisions at the Future Circular Collider\\}
\vspace{0.5cm}
\Author{David d'Enterria\davidemail}
\Address{\david}
\Author{\underline{Patricia Rebello Teles}\patemail}
\Address{\patricia}
\Author{Daniel E. Martins\danielemail}
\Address{\daniel}

\vspace{0.3cm}
\begin{Abstract}
\noindent
The measurements of the two-photon production of the Higgs boson and of $\rm W^\pm$ boson pairs 
in $\ee$ collisions at the Future Circular Collider (FCC-ee) are investigated. The processes 
$\ee\xrightarrow{\gaga}e^+\,{\rm H}\,e^-,e^+\,{\rm W^+W^-}\,e^-$ are computed using the effective photon 
approximation for electron-positron beams, and studied in their ${\rm H}\to\bbbar$ and ${\rm W^+W^-}\to 4j$ 
decay final-states including parton showering and hadronization, jet reconstruction, $e^\pm$ forward
tagging, and realistic experimental cuts. After selection criteria, up to 75 Higgs bosons and 6600 $\rm W^{\pm}$ 
pairs will be reconstructed on top of controllable continuum backgrounds at $\sqrt{s} = $~240 and 350~GeV 
for the total expected integrated luminosities, by tagging the scattered $e^\pm$ with near-beam detectors. 
A 5$\sigma$ observation of $\gaga \to$~H is thereby warranted, as well as high-statistics studies of
triple  $\rm \gamma WW$ and quartic $\rm \gamma\gamma WW$ electroweak couplings, improving by at least 
factors of 2 and 10 the current limits on dimension-6 anomalous quartic gauge couplings.
\end{Abstract}
\vfill
\begin{Presented}
EDS Blois 2017 Conference , Prague, \\ Czech Republic, June 26--30, 2017
\end{Presented}
\vfill
\end{titlepage}
\def\thefootnote{\fnsymbol{footnote}}
\setcounter{footnote}{0}
%

\section{Introduction}

\begin{figure}[htbp]
  \centering
  \includegraphics[width=10pc]{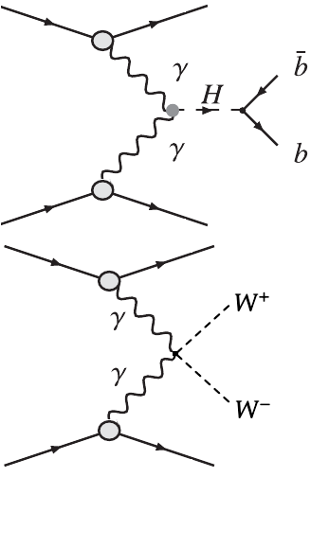}
  \includegraphics[width=25pc]{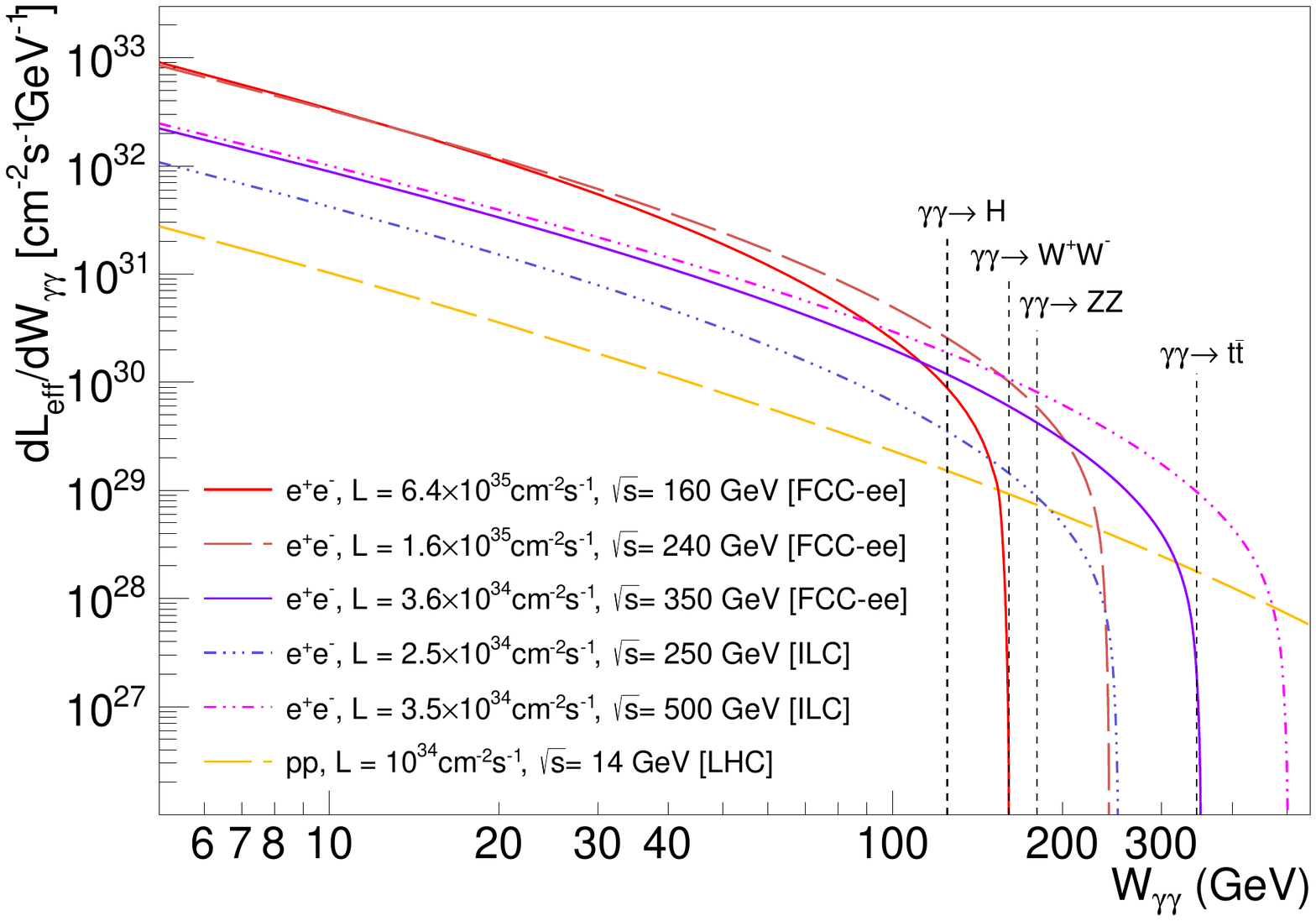}
  \caption{\label{fig1} {\small Left: Diagrams for the two-photon production of the Higgs boson (top) and W$^\pm$ pairs (bottom).
Right: Two-photon effective luminosities (EPA fluxes) as a function of $\gaga$ \cm\ energy over $W_\gaga \approx$~5--400~GeV in $\ee$ collisions
at FCC-ee and ILC, and in pp collisions at the LHC.}}
  \end{figure}

After the Higgs boson discovery at the LHC~\cite{Chatrchyan:2012xdj,Aad:2012tfa}, subpercent precision studies of its couplings 
to all Standard Model (SM) particles, sensitive to scalar-coupled new physics in the multi-TeV range~\cite{dEnterria:2017dac}, 
require an electron-positron ``Higgs factory"  with high luminosities and low backgrounds running at center-of-mass (\cm)
energies of 
$\sqrts\approx$~240--350~GeV. Such conditions can be met in the $\ee$ mode of the Future Circular Collider 
(FCC-ee)~\cite{tlep}, a post-LHC project under consideration at CERN, based on a 80-km circular ring aimed at eventually 
running proton-proton collisions up to \cm\ energies of $\sqrts = 100$~TeV~\cite{Mangano:2016jyj}.
Among many important SM and beyond SM channels~\cite{tlep,dEnterria:2016sca}, the large FCC-ee luminosities would make it feasible
for the first time to observe with high rates the production of high-mass systems in photon-photon collisions thanks to the 
large effective fluxes of quasireal $\gamma$'s~\cite{epa} radiated from the high-luminosity $\ee$ beams~\cite{Teles:2015xua}. 
The two-photon production of the Higgs boson and of W$^\pm$ pairs (Fig.~\ref{fig1}, left) are both accessible 
at the FCC-ee and provide interesting tests of the SM electroweak sector. 
The former process provides an independent measurement of the H-$\gamma$ coupling not based on Higgs decays but on its
$s$-channel production mode, whereas the latter process probes trilinear $\rm \gamma W^{+}W^{-}$ 
and quartic $\rm \gaga W^{+}W^{-}$ electroweak couplings, and allows in particular to place competitive
limits on anomalous quartic gauge couplings (aQGC) parametrized with dimension--6 and 8 effective operators of 
new physics at higher energy scales $\Lambda$, as done in pp collisions at the LHC~\cite{CMSaaww}.
Figure~\ref{fig1} (right) shows the $\gaga$ effective luminosities ($\Lumi_{_{\rm eff}}$) as a function of photon-photon \cm\
energy $W_\gaga$, obtained from the convolution of the corresponding effective photon approximation (EPA)
fluxes~\cite{epa} with the collider luminosities at various planned $\ee$ colliders (FCC-ee, ILC) and in 
pp collisions at the LHC. Except for electromagnetic (ultraperipheral) ion-ion collisions at the FCC~\cite{Dainese:2016gch},  
the FCC-ee features the largest effective two-photon luminosities for masses $W_\gaga\lesssim$~200~GeV,
with the advantage of the absence of pile-up collisions.\\

In this work, feasibility studies for the measurement of photon-fusion production of the Higgs boson 
in its $b\bar{b}$ decay channel, and of W$^\pm$ boson pairs decaying into four jets are presented.
Observing both fully-hadronic exclusive final-states is unfeasible at the LHC due to the large pp pileup 
conditions and the lack of (420~m) proton taggers in the tunnel with a good acceptance in the ${\cal O}\rm (100~GeV)$ 
mass region~\cite{fp420}. Compared to our previous studies~\cite{Teles:2015xua}, several improvements have been implemented.
First, instead of using \madgraph~5~\cite{madgraph} 
the events are now generated with the latest versions of \pythia~8~\cite{pythia8} and \superchic~v2~\cite{superchic}, 
keeping the exact kinematics information of the scattered $e^\pm$;
showering and hadronization of the final-state quarks are taken care by \pythia~6 or 8; 
and jets are reconstructed with \fastjet. Second, the $e^\pm$ tagging conditions have been changed. 
Previously, the requirement to observe the scattered $e^\pm$ within the instrumented central
detector reduced significantly the visible cross sections. Our approach now, is to tag the outgoing
$e^\pm$ in near-beam detectors inside the FCC-ee tunnel, as done at the LHC for the proton-proton case~\cite{ctpps}.
Requiring double $e^\pm$ tagging within 1$^\circ$ of the beam line, increases significantly the acceptance for 
both $\gaga$ processes: 75\%, 95\% and 98.5\% for Higgs production at $\sqrts$~=~160, 240, and 350~GeV respectively, although
a realistic study of the FCC-ee beam optics is needed in order to determine the exact position of such
$e^\pm$-taggers in the tunnel.



\section{Theoretical setup}

Event generation for the two-photon signal processes $\ee\xrightarrow{\gaga}e^+\,{\rm H}(\bbbar)e^-;\,e^+\,{\rm W^+W^-}(4j)\,e^-$,
(as well as the corresponding $\gaga$ backgrounds) is carried out with \superchic~2.04~\cite{superchic} and/or \pythia~8.226~\cite{pythia8} 
Monte Carlo (MC) codes, where the cross sections are obtained from the convolution of 
the corresponding EPA photon fluxes with the $\rm \gamma H,\;\gamma WW,\;\gamma\gamma WW$ vertices described by matrix elements 
(effective ones, in the $\gamma$H case) and exact kinematics. The virtuality of both photons is constrained to be in the quasireal
range, $Q^{2} < 2\;\mbox{GeV}^{2}$. Both MC event generators yield consistent cross sections for the same processes. Non-photon-induced
backgrounds from $\ee$ collisions sharing the same final-states as the signals of interest are generated with \pythia~6.4~\cite{pythia6} or \pythia~8.226.
Initial and final state radiation (ISR, FSR), parton showering, hadronization, and decays are handled also by any of the two
\pythia\ codes. The jets of hadrons resulting from the fragmentation of the produced partons 
are reconstructed using the $e^{+}e^{-}\;k_{t}$ Durham algorithm~\cite{kTalgo} with the \fastjet~3.0 library~\cite{fastjet}.



\section{$\gaga \to {\rm H} \to b \bar{b}$ results}

\begin{figure}[h]
\centering
\includegraphics[width=8.3cm]{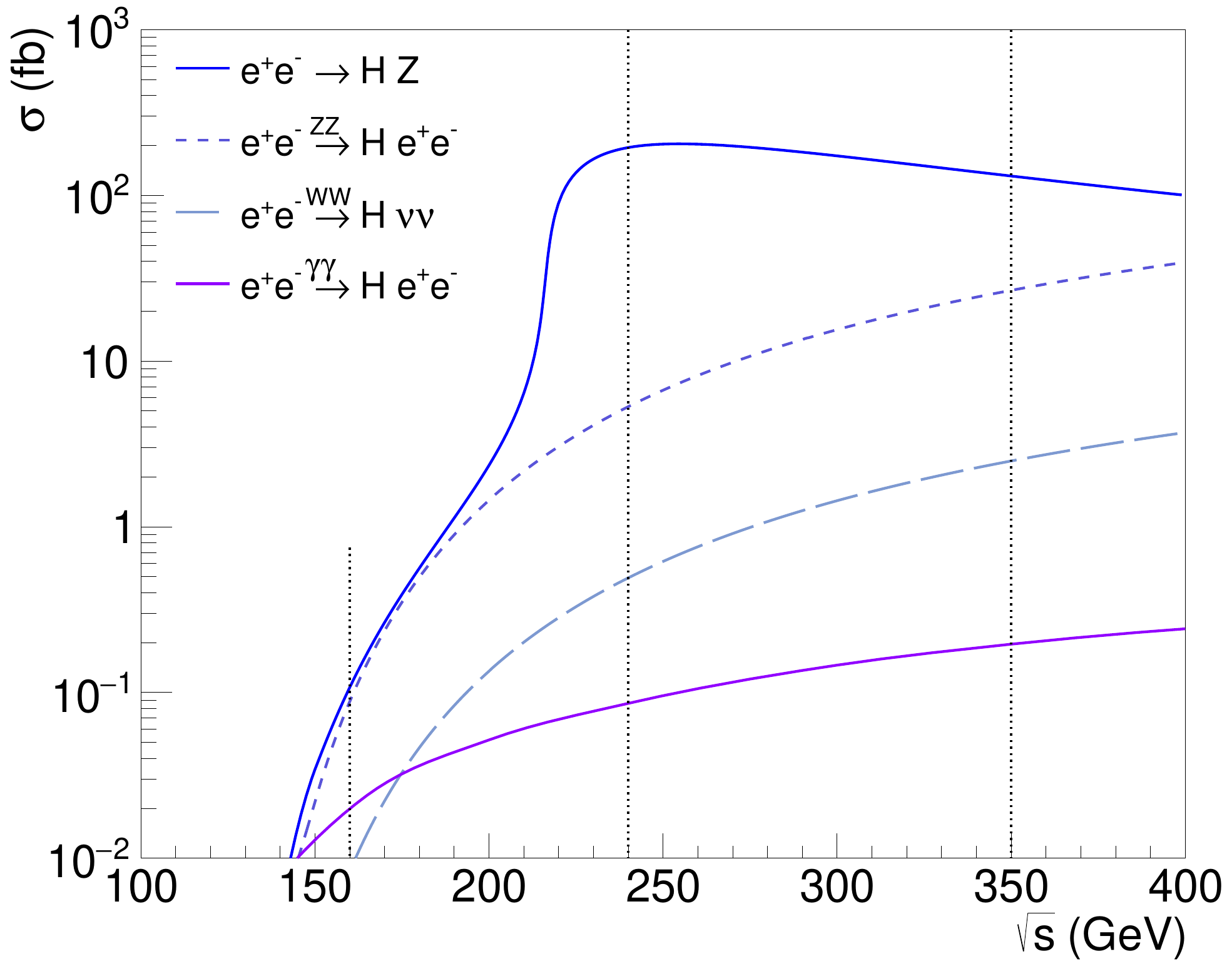}
\includegraphics[width=8.0cm]{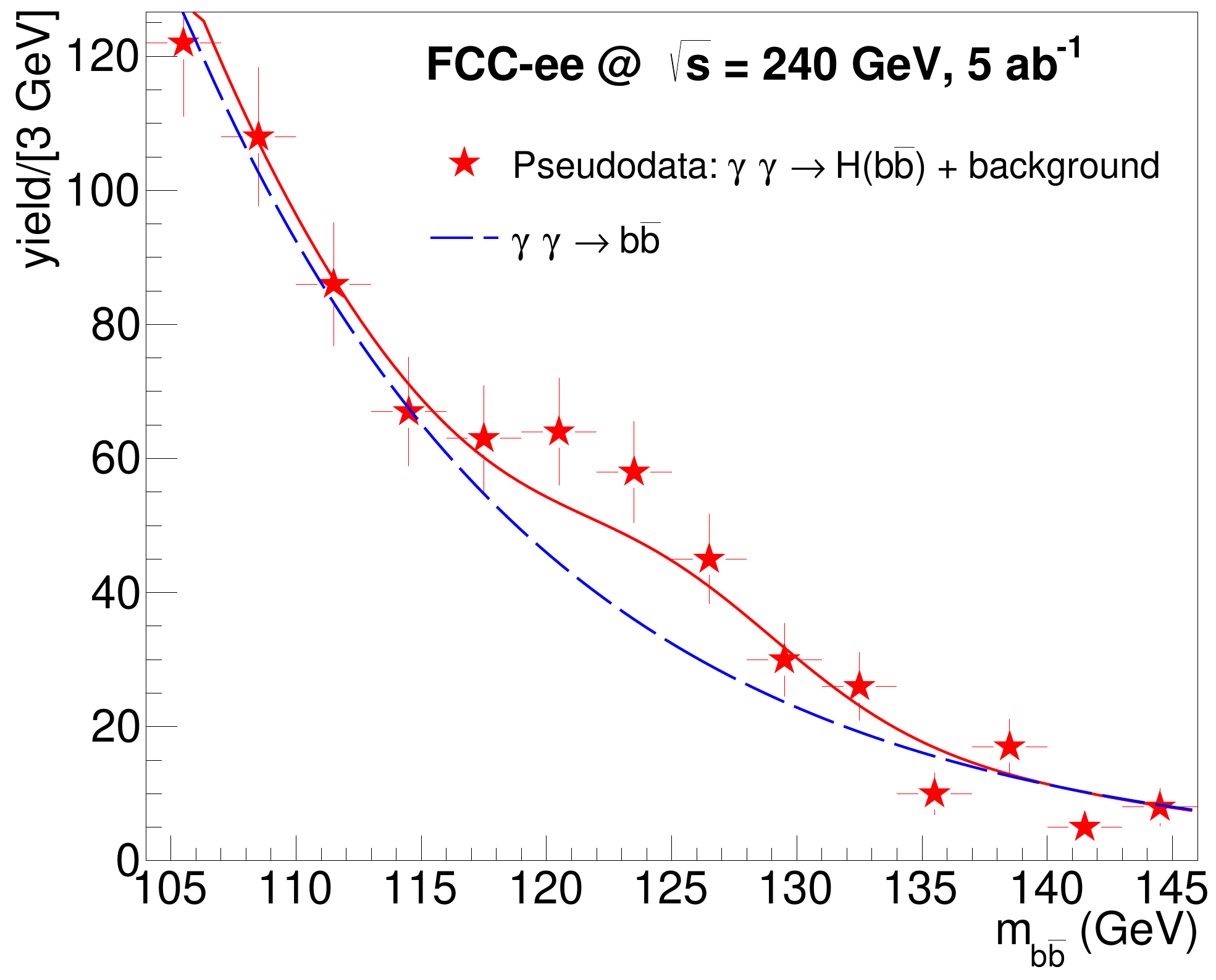}
\caption{\label{fig:sigmaH_vs_sqrts} {\small Left: Contributions to the Higgs boson cross section
in $\ee$ collisions (unpolarized beams, ISR included) as a function of \cm\ energy: HZ and VBF 
(via ZZ and WW exchanges) computed with {\sc hzha}~\cite{HZHA}, and photon-fusion computed with
\superchic~2.04~\cite{superchic}. The vertical lines indicate the FCC-ee energies 
of $\sqrts$~=~160, 240, and 350~GeV. Right: Expected dijet invariant mass 
distribution for two-photon Higgs signal and $\bbbar$ continuum, in 5~ab$^{-1}$ of 
$\ee$ collisions at $\sqrts$~=~240~GeV, after $e^\pm$ tagging and simple cuts in the b-jet pseudorapidities (see text).}}
\end{figure}

Figure~\ref{fig:sigmaH_vs_sqrts} (left) shows the energy-dependence of the  cross sections for the ``standard''  
Higgsstrahlung and W,Z boson fusion (VBF) production mechanisms computed with {\sc hzha}~\cite{HZHA},
compared to $\gaga$ production (bottom curve) in the range of FCC-ee energies. 
Two-photon Higgs production has the smallest cross sections, amounting
to about 25, 90, and 200~ab at $\sqrt{s}=$~160, 240, and 350~GeV respectively, which would still lead
to the production of 130--300 scalar bosons thanks to the large total integrated luminosities 
expected at each \cm\ energy ($\LumiInt\approx$~10, 5, 1~ab$^{-1}$ respectively). 
Of course, the visible number of Higgs bosons is smaller after accounting for decay branching ratios, reconstruction performance, and analysis cuts 
to reduce backgrounds. The dominant Higgs decay mode is $\rm H\to\bbbar$, with 58\% branching fraction~\cite{hdecay},
and thereby the channel with the largest number of counts expected.
Signal and background events for the $\gaga \to {\rm H} \to b \bar{b}$ measurement are 
simulated with \pythia~8.226, with \fastjet~3.0 being used to reconstruct two exclusive jets final-states. 
Reducible backgrounds, dominated by the process $e^{+}e^{-}\to Z^*/\gamma^*\to b\bar{b}$ with a cross section 
of $\sigma \approx$~2~pb over the mass range $m_{\bbbar}$~=~100--150~GeV, can be completely 
removed by requiring the double $e^{\pm}$ tagging at polar angles $\theta_{e^\pm} < 1^{\circ}$. 
At 160, 240, and 350~GeV, double-tagging saves 75\%, 95\%, and 98.5\% of the two-photon Higgs signal, 
while it also completely gets rid of 
the $\ee\xrightarrow{ZZ}e^+\,{\rm H}\,e^-$ process (second dominant curve in 
Fig.~\ref{fig:sigmaH_vs_sqrts}, left) which features outgoing $e^\pm$ at much central rapidities.
The irreducible background from the $\gaga \to b\bar{b}$ continuum is 30--40 times larger than the 
signal over masses $100 < W_{\gaga} < 150$~GeV, but can be suppressed (as well as that from misidentified 
$c\bar{c}$ and $q\bar{q}$ pairs) via various kinematical cuts.
The data analysis follows the similar approach described in~\cite{dEnterria:2009cwl}.
The following reconstruction performances have been assumed: jets reconstructed over $|\eta|<5$,
7\% $b$-jet energy resolution (resulting in a dijet mass resolution of $\sim$6~GeV at the Higgs peak), 
75\% $b$-jet tagging efficiency, and 5\% (1.5\%) $b$-jet mistagging probability for a $c$ (light-flavour $q=udsg$) 
quark. For the double $b$-jet final-state of interest, these lead to a $\sim$56\% efficiency for the pure generated signal (S), 
and a total reduction of the misidentified $c\bar{c}$ and $q\bar{q}$ continuum backgrounds (B) by factors of 
$\sim$400 and $\sim$400\,000, respectively (Table~\ref{tab:1}).
\begin{table}[h]
\caption{\label{tab:1} \small Summary of the visible cross sections for signal and backgrounds in the 
$\gaga \to H(\bbbar)$ analysis, obtained from \pythia~8 simulations in $\ee$  collisions at 
$\sqrts$~=~160,~240~GeV and~350~GeV, after applying various selection criteria.}
\vspace{0.5cm} 
\centering
\small{\begin{tabular}{l|c|c|c|c}\hline
Process &$\sqrt{s}$  & $[m_{jj}=$~100--150~GeV]  &  $\cdots$   &  $\cdots$  \\
        & (GeV)      & ($b$-jet (mis)tag efficiency)     &  $\cdots$     & $\cdots$      \\
        &            &                                   & $e^{\pm}$-tag &  $\cdots$ \\
        &            &                                   & $|\eta^j|<1$  &  $\cdots$ \\
        &            &                                   &               &  $[m_{jj}=$~117--133~GeV]\\\hline
$\gaga \to H \to b\bar{b}$  & 160 & 13.2 (7.4) ab    & 4.0 ab  & 3.3 ab  \\   
$\gaga \to b\bar{b}$        &     & 547. (308.) ab   & 41. ab  & 13.5 ab \\ 
$\gaga \to c\bar{c}$        &     & 13.2 fb (33. ab) & 3.0 ab  & 0.92 ab \\
$\gaga \to q\bar{q}$        &     & 18.5 fb (4.2 ab) & 0.34 ab & 0.12 ab \\
\hline
$\gaga \to H \to b\bar{b}$ & 240  & 58.0 (32.6) ab   & 17.7 ab  & 14.9 ab  \\ 
$\gaga \to b\bar{b}$        &     & 2.05 (1.15) fb   & 125. ab  & 52.2  ab \\
$\gaga \to c\bar{c}$        &     & 4.95 fb (124 ab) & 9.2 ab   & 3.71 ab   \\
$\gaga \to q\bar{q}$        &     & 69.5 fb (15.6 ab) & 0.94 ab & 0.37 ab  \\
\hline
$\gaga \to H \to b\bar{b}$ & 350  & 130. (73.1) ab  & 30.3 ab  & 25.5  ab \\
$\gaga \to b\bar{b}$        &     & 4.38 (2.47) ab  & 204. ab  & 96.7 ab \\
$\gaga \to c\bar{c}$        &     & 106. fb (264. ab) & 14.6 ab & 6.92  ab \\
$\gaga \to q\bar{q}$        &     & 147. fb (33.1 ab) & 1.5 ab  & 1.06 ab \\
\hline
\end{tabular}
}
\end{table}

Various simple kinematical cuts can be applied to enhance the S/B ratio.
The Higgs boson is produced in the $s$-channel and its associated decay $b$-jets, emitted isotropically, 
have pseudorapidities peaking around $\eta^j\approx 0$, whereas the continuum -- with quarks propagating 
in the $t$- or $u$- channels -- are more peaked in the forward and backward directions. Simply requiring both jets 
to have pseudorapidities $|\eta^j|<1$ reduces the signal by 30--50\%, while removing 80--90\% of the backgrounds. 
The significance of the signal can be computed from the resulting number of counts within 1.4$\sigma$ around the
Gaussian Higgs peak (\ie\ $117 < m_{\bbbar} < 133$~GeV) over the underlying exponential dijet continuum.
At $\sqrts$~=~240 GeV, for a total integrated luminosity of $\LumiInt = 5\;\mbox{ab}^{-1}$ (3 years, 2 interaction points), 
we expect about 75 signal counts over 275 for the sum of backgrounds, reaching a statistical significance 
close to 5$\sigma$ (Fig.~\ref{fig:sigmaH_vs_sqrts}, right). Similar estimates for 160~GeV (10~ab$^{-1}$) 
and 350~GeV (1~ab$^{-1}$) yield 3$\sigma$ significances for the evidence of $\gaga\to$~H production.
Of course, those numbers are based on a simple set of kinematical cuts. A multivariate analysis
exploiting many other kinematical properties of signal and backgrounds would easily improve such results.

\section{$\rm \gaga \to W^{+}W^{-} \to 4j$ results}

\begin{figure}[htbp!]
  \centering
\includegraphics[width=5.4cm]{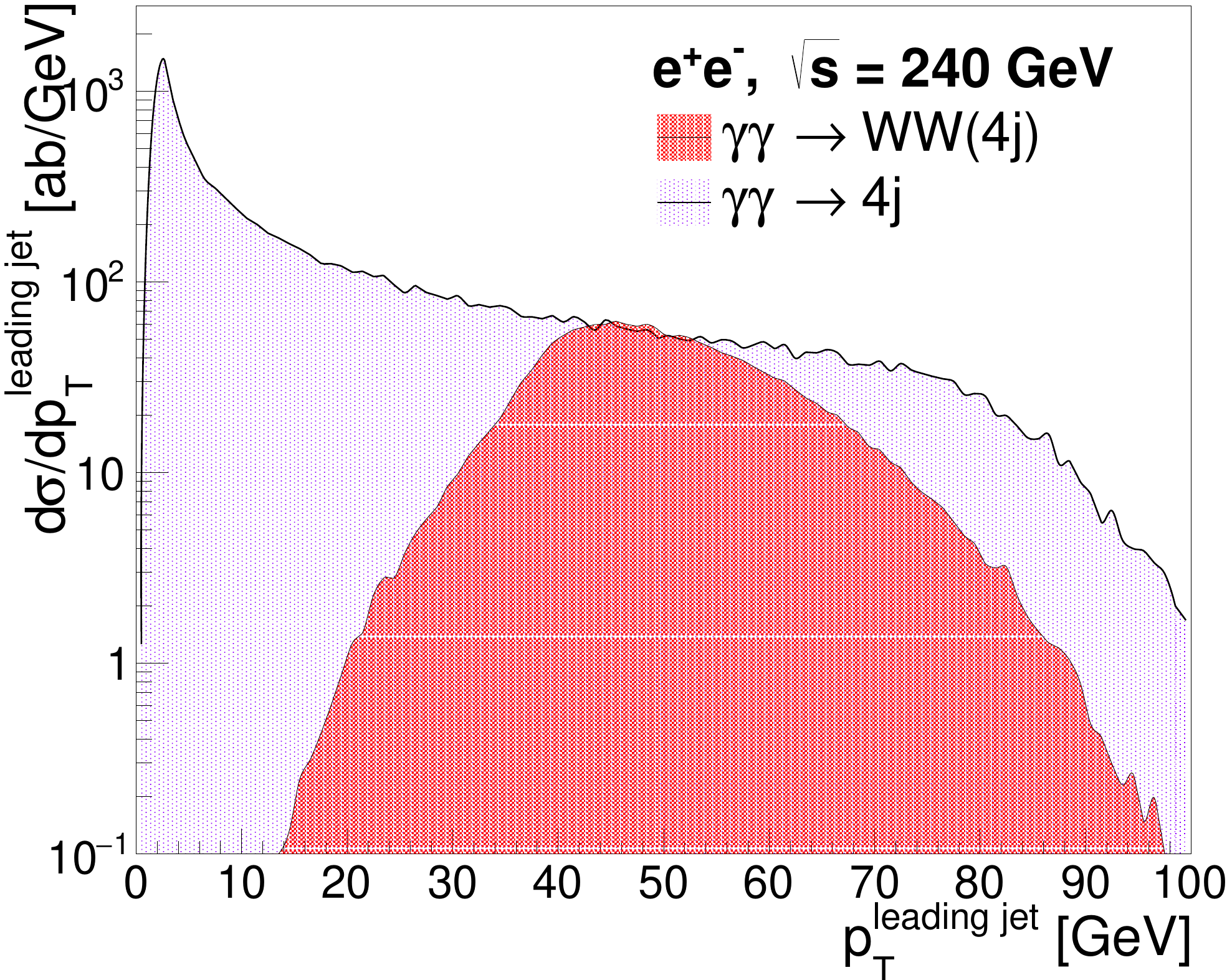}
\includegraphics[width=5.4cm]{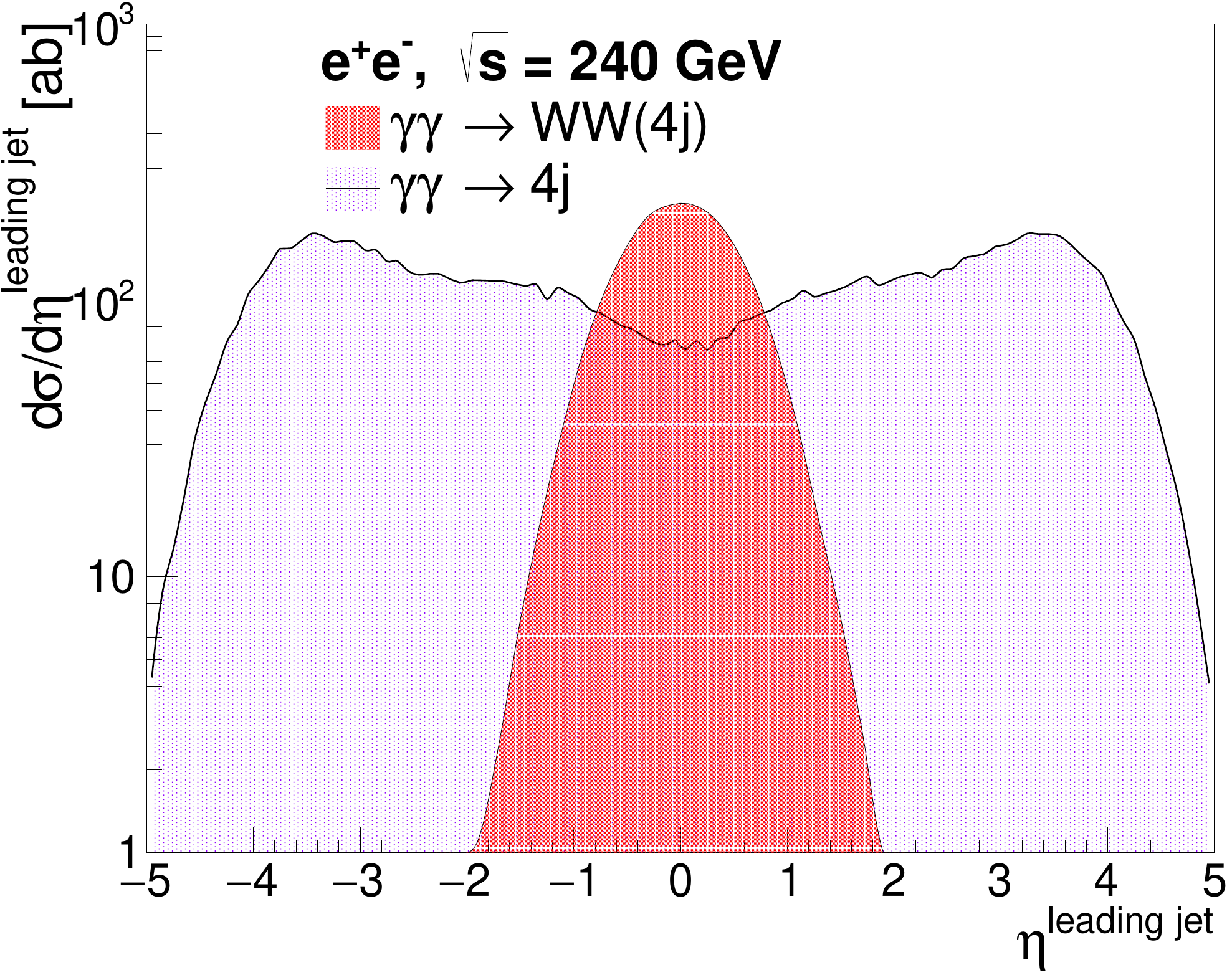}
\includegraphics[width=5.4cm]{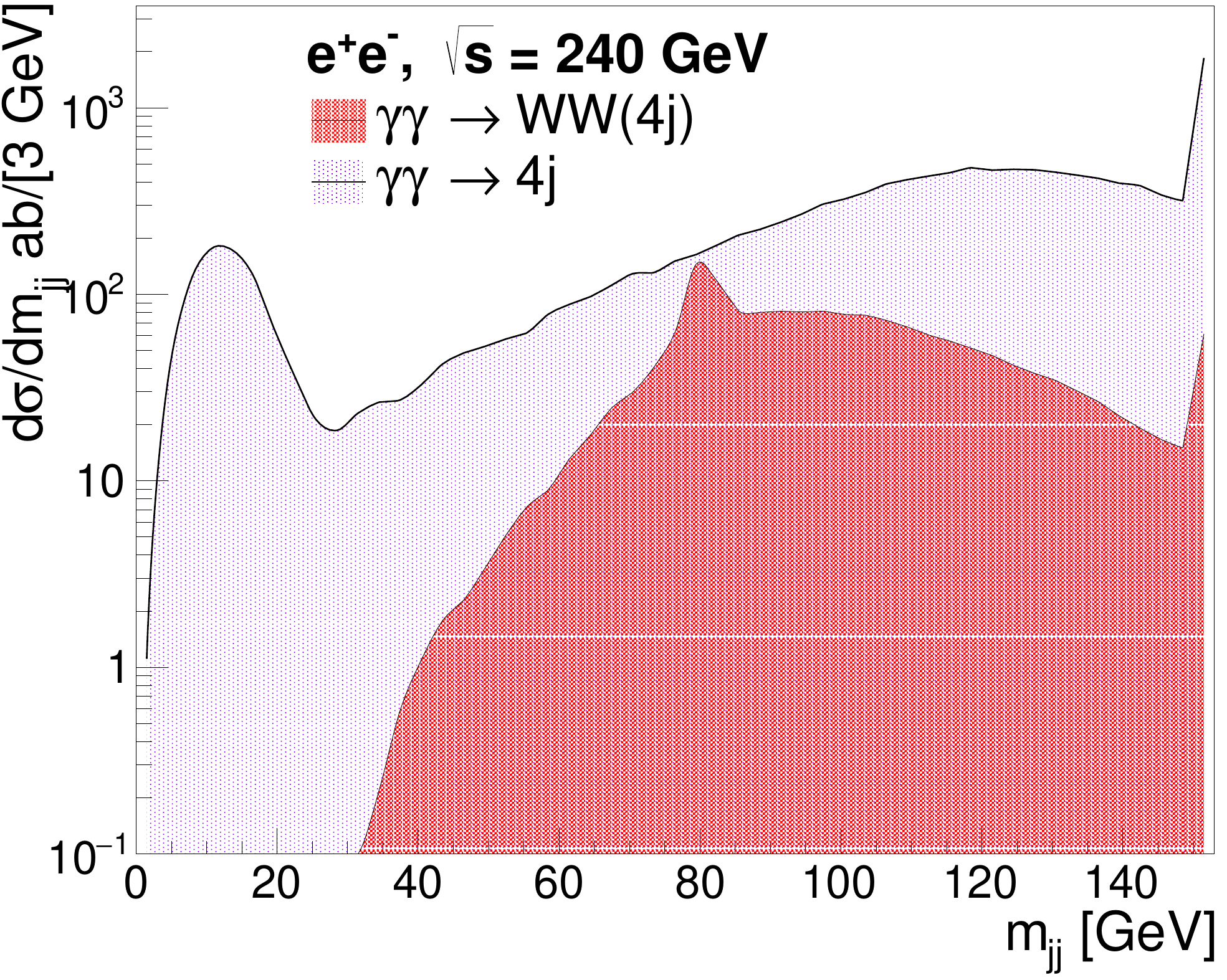}
\includegraphics[width=5.4cm]{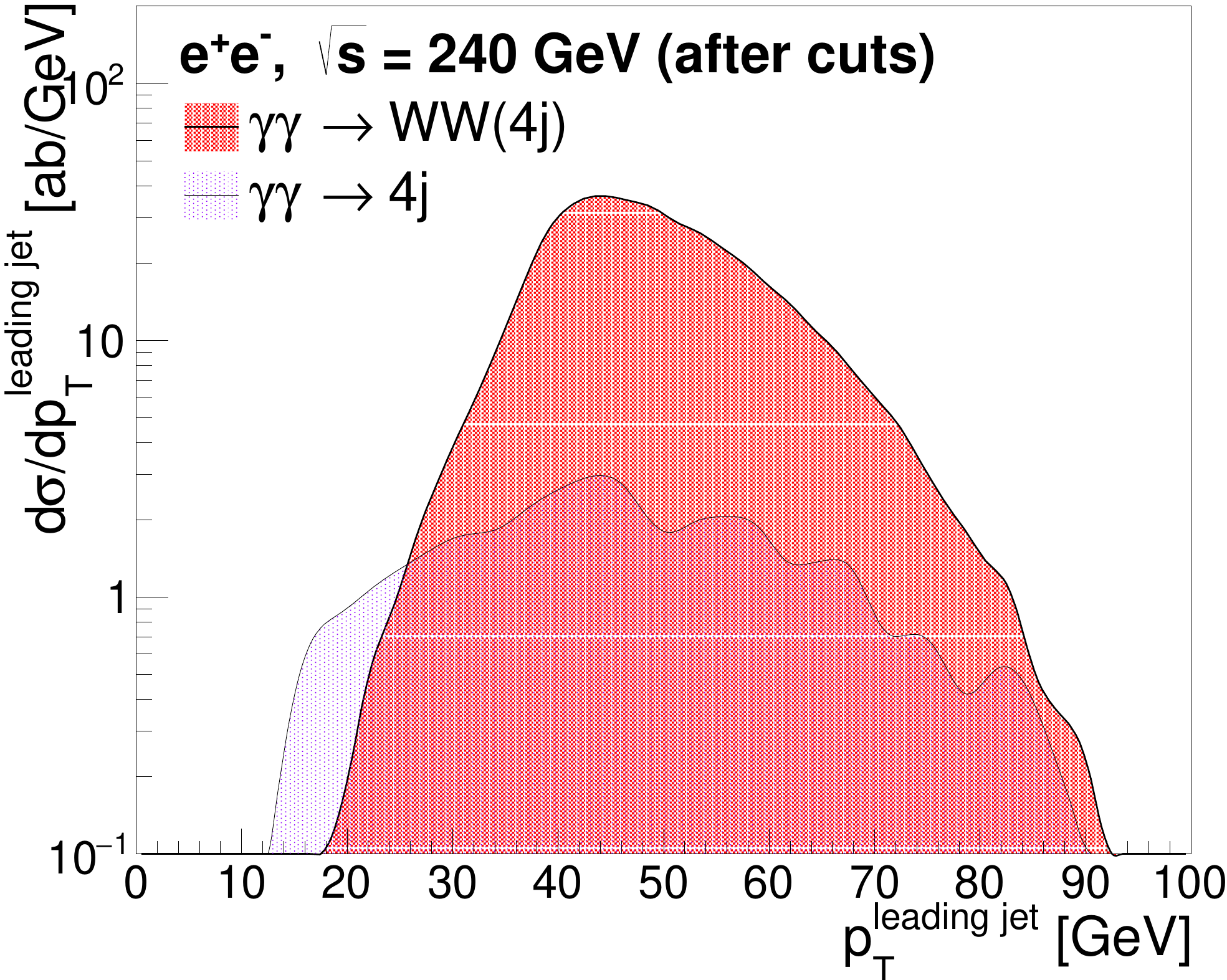}
\includegraphics[width=5.4cm]{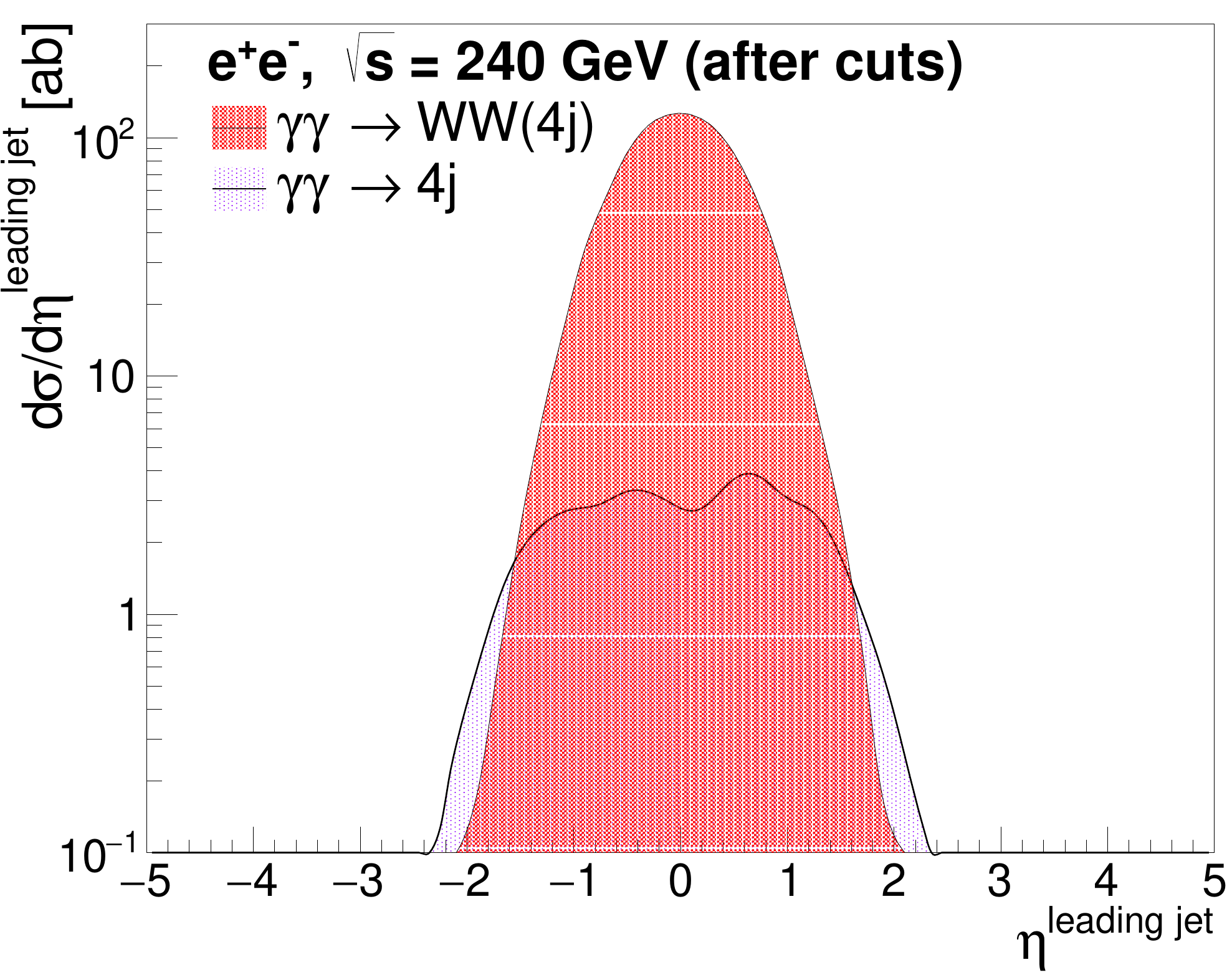}
\includegraphics[width=5.4cm]{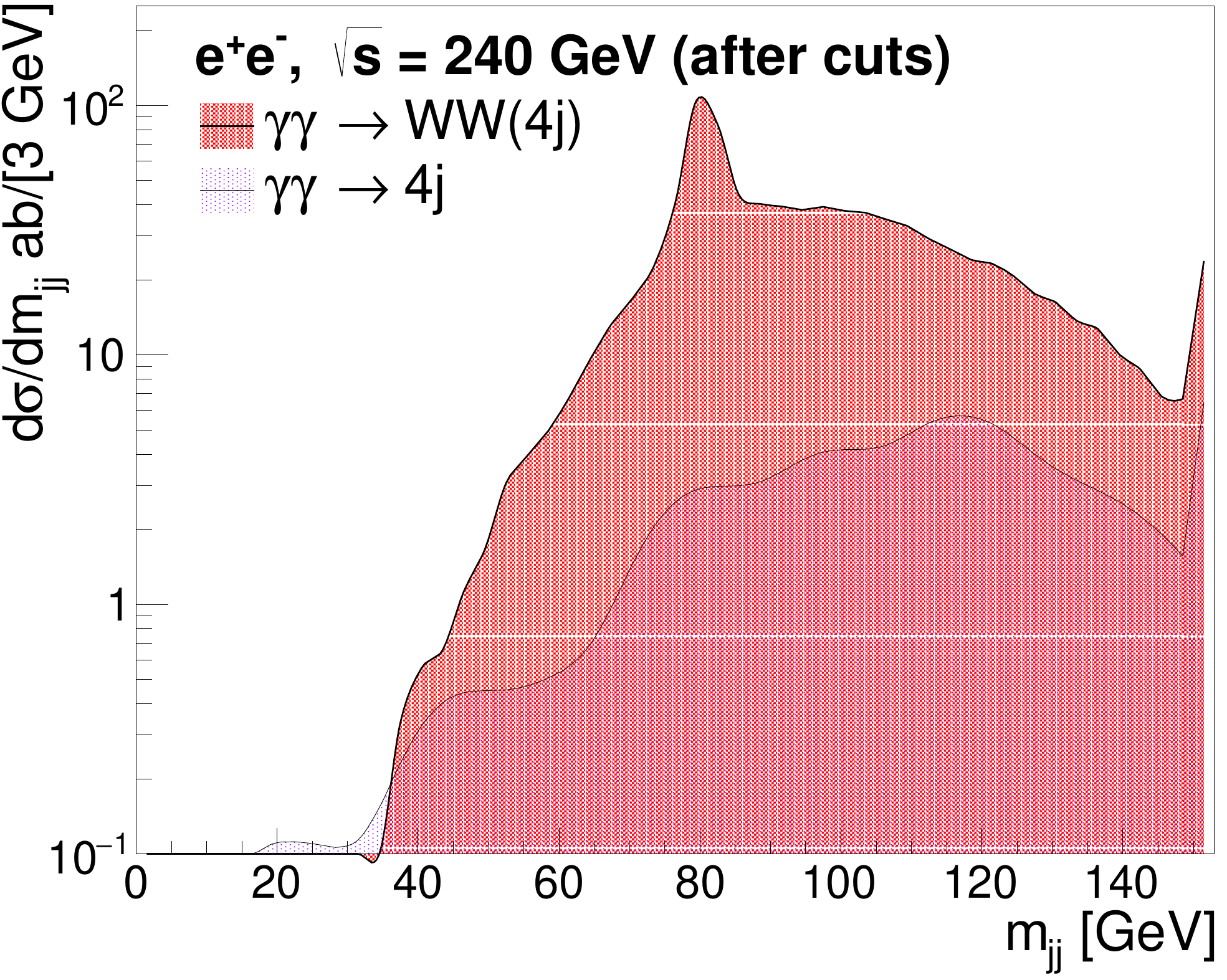}
\caption{\label{talk240GeV}{\small Relevant kinematical distributions for $\rm \gaga\to WW$ signal and 
$\gaga\to 4j$ background in $\ee$ collisions at $\sqrts = 240$~GeV: 
$p_T$ (left) and $\eta$ (middle) of the leading jet, and invariant mass $m_{jj}$ of jet pairs (right),
before (top) and after (bottom) applying kinematical cuts.}}
\end{figure}

The \superchic~v2.04 code is used for event generation of $\rm W^{+}W^{-}$ pairs in $\ee$ at $\sqrts = 240, 350$~GeV
in the phase space region 
given by $W_{\gaga} > 161$~GeV. The two bosons, decayed, showered, and hadronized with \pythia~8.226,
are reconstructed as a four exclusive jets final-state. The dominant background from the process 
$e^{+}e^{-}\to 4j$ with $\sigma\approx$~8.8~pb, is fully suppressed with double $e^{\pm}$ tagging
($\theta_{e^{\pm}}<1$), leaving as irreducible background only the $\gaga \to 4j$ continuum process 
(simulated with \pythia~6.4) which can be easily removed applying a few simple kinematical cuts (Table~\ref{tableaaww}).
Requiring all jets to have $p_{T} > 10$~GeV, $|\eta^{j}| < 2.5$, and be separated by $\Delta R^{j} > 0.4$ 
reduces by $\sim$97\% the background for a $\sim$25\% loss of the signal. 
Figure~\ref{talk240GeV} compares the relevant differential distributions for signal and irreducible background
in $\ee$ at $\sqrts =$~240~GeV before (top) and after (bottom) applying the selection criteria. Requiring, in addition,
each pair of jets to have an invariant mass around the W value ($m_{jj}=$ 76.5--84.5~GeV) further improves the signal over
background (last column of Table~\ref{tableaaww}). 
At $\sqrts = 240, 350$~GeV with total integrated luminosities of $\LumiInt\approx$~5 and 1~ab$^{-1}$, 
we would therefore expect about 4400 and 6600 signal events, respectively, over much smaller backgrounds. 

\begin{table}[htbp!]
\caption{\label{tableaaww} Summary of the visible cross sections for signal and backgrounds in the 
$\rm \gaga\to W^+W^- \to$~4-jets analysis, obtained with \superchic~2.04 (plus \pythia~8) and \pythia~6 
simulations in $\ee$ at  $\sqrts$~=~240~GeV and 350 GeV, after applying various selection criteria.}
\vspace{0.5cm} 
\centering
\small{\begin{tabular}{lcccc}
\hline
Process & $\sqrt{s}$  & cross section         & $\cdots$                     &  $\cdots$  \\
        &             & (incl.   $e^{\pm}$-tag) & $p_{T}^{j} > 10 \mbox{GeV}$  &  $\cdots$  \\
        &             &                         & $|\eta^{j}| < 2.5$           &  $\cdots$  \\
        &             &                         & $\Delta R^{jj} > 0.4$        &  $\cdots$  \\
        &             &                         &                        & $[m_{jj}=$ 76.5--84.5~GeV] \\
\hline
$\rm\gamma \gamma \to W^{+}W^{-} \to 4 \mbox{jets}$  & 240 GeV  & 1.64 (1.43) fb   & 1.10 fb & 0.87 fb \\
$\rm\gamma \gamma \to 4 \mbox{jets}$                 &      & 10.49 (9.12) fb  & 0.27 fb & 0.10 fb \\
\hline
$\rm\gamma \gamma \to W^{+}W^{-} \to 4 \mbox{jets}$  & 350 GeV & 10.5 (10.2) fb   &  7.7 fb & 6.61 fb \\
$\rm\gamma \gamma \to 4 \mbox{jets}$                 &      & 46.9 (45.4) fb   &  1.3 fb & 0.52 fb \\
\hline
\end{tabular}
}
\end{table}

\begin{figure}[htbp!]
  \centering
\includegraphics[width=8.cm,height=7.cm]{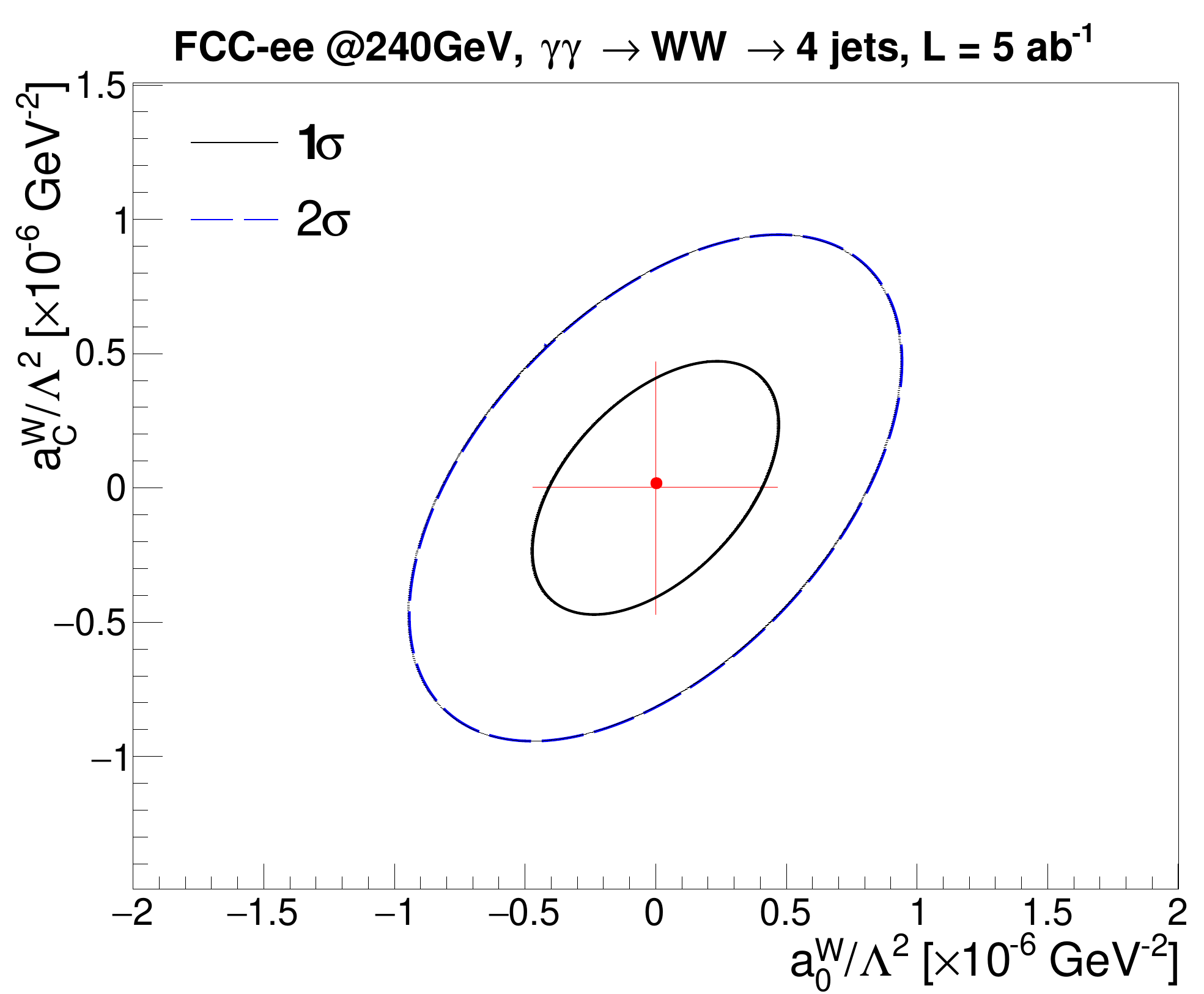}
\caption{\label{fig:aQGC}{\small Expected 1$\sigma$ and 2$\sigma$ limits for the anomalous quartic gauge coupling parameters 
$a_0^W/\Lambda$ and $a_c^W/\Lambda$, from the $\ee\xrightarrow{\gaga}{\rm W^+W^-}(4j)$ 
measurement at $\sqrts$~=~240~GeV (FCC-ee, $\LumiInt =$~5~ab$^{-1}$).}}
\end{figure}

A preliminary study based on an implementation of dimension-6 $\rm \gaga WW$ operators in \madgraph~5 (v2.5.4)~\cite{madgraph}, 
indicates that using the $p_T$ of the leading jet and the invariant mass of the WW system as discriminating variables, 
from the expected number of events at 240~GeV we can forecast at least factors of 2 and 10 improvements on the limits
of the aQGC parameters $a_0^W/\Lambda^{2}$ and $a_0^c/\Lambda^{2}$ : $(-0.47;+0.47)\times 10^{-6}$~GeV$^{-2}$
(Fig.~\ref{fig:aQGC}), in comparison with the latest results derived (without form-factor) in pp collisions at the LHC~\cite{CMSaaww}.

\section{Conclusions}

Feasibility studies have been presented for the measurements of the two-photon production of the Higgs boson 
(in the $b\bar{b}$ decay channel) as well as of $\rm W^{+}W^{-}$ pairs (in their fully-hadronic decay mode) 
in $\ee$ collisions at the FCC-ee, using the equivalent photon flux of the colliding beams. 
Both final-states are inaccessible at the LHC due to the huge pileup and QCD backgrounds in their  
full-jet decay channels. Results have been presented for collisions at center-of-mass energies of 
$\sqrt{s} = 160$, 240, and 350~GeV using \superchic\ and \pythia~8 Monte Carlo simulations based on the EPA
approach, including parton showering, hadronization, and exclusive (2 or 4) jet reconstruction with the $k_T$
Durham algorithm. Realistic jet reconstruction performances, and (mis)tagging efficiencies are considered, as well as 
kinematical selection criteria to enhance the signals over the relevant backgrounds. By tagging the outgoing $e^\pm$ 
with near-beam detectors ($\theta_e<1^\circ$), the two-photon $s$-channel
production of the Higgs boson can be observed with 5~ab$^{-1}$ integrated at $\sqrts$~=~240~GeV, 
where we expect about 75 signal counts on top of 275 counts for the sum of $\gaga$ dijet backgrounds.
The measurement of $\rm \gaga \to WW \to 4$~jets will yield several thousands signal events after cuts, that will allow for 
detailed studies of the trilinear $\rm \gamma WW$ and quartic $\rm \gaga WW$ couplings, improving by at least factors of 
2 and 10 the current limits on dimension-6 anomalous quartic gauge couplings. The feasibility analyses developed in this work 
confirm the novel Higgs and electroweak physics potential open to study through $\gaga$ collisions at the FCC-ee.


\subsection*{Acknowledgments}
We thank Lucian Harland-Lang and Ilkka Helenius for useful discussions on \superchic\ and \pythia~8 respectively.
P.R.T acknowledges support from the EDS Blois 2017 organizers and Michelangelo Mangano.
This work has granted partially by the Coordination for the Improvement of Higher Education Personnel -- CAPES (Brazil).

\end{document}